\documentclass[twocolumn,english,prl,showpacs]{revtex4}
\usepackage[T1]{fontenc}
\usepackage[latin9]{inputenc}
\usepackage{textcomp}
\usepackage{amsmath}
\usepackage{amssymb}
\usepackage{graphicx}
\usepackage{esint}

\makeatletter
\@ifundefined{textcolor}{}
{%
 \definecolor{BLACK}{gray}{0}
 \definecolor{WHITE}{gray}{1}
 \definecolor{RED}{rgb}{1,0,0}
 \definecolor{GREEN}{rgb}{0,1,0}
 \definecolor{BLUE}{rgb}{0,0,1}
 \definecolor{CYAN}{cmyk}{1,0,0,0}
 \definecolor{MAGENTA}{cmyk}{0,1,0,0}
 \definecolor{YELLOW}{cmyk}{0,0,1,0}
}

\makeatother

\usepackage{babel}
\begin{document}

\title{Efficient narrowband teraherz radiation \\ from electrostatic wakefields
in non-uniform plasmas}

\author{Alexander Pukhov}

\affiliation{Institut fuer Theoretische Physik I, Universitaet Duesseldorf, 40225
Germany}

\author{Anton Golovanov and Igor Kostyukov}

\affiliation{IAP RAS, Nizhni Novgorod, Russia}
\begin{abstract}
It is shown that electrostatic plasma wakefields can efficiently radiate
at harmonics of the plasma frequency when the plasma has a positive
density gradient along the propagation direction of a driver. The
driver propagating at a sub-luminal group velocity excites the plasma
wakefield with the same phase velocity. However, due to the positive
density gradient, the wake phase velocity steadily increases behind
the driver. As soon as the phase velocity becomes super-luminal, the
electrostatic wakefield couples efficiently to radiative electromagnetic
modes. The period of time when the phase velocity stays above the
speed of light depends on the density gradient scale length. The wake
radiates at well-defined harmonics of the plasma frequency in the
teraherz (THz) band. The angle of emission depends on the gradient
scale and the time passed behind the driver. For appropriate plasma
and driver parameters, the wake can radiate away nearly all its energy,
which potentially results in an efficient, narrow band and tunable
source of THz radiation.
\end{abstract}

\pacs{PACS1}

\maketitle
The teraherz (THz) band of electromagnetic radiation spans the frequency
range $3\times10^{11}{\rm Hz}$ to $3\times10^{13}{\rm Hz}$ \cite{Sirtori GAP}.
From the engineering point of view, there are currently few practical
radiation sources in this THz gap \cite{GAP devices}. The standart
vacuum devices (gyrotrones, magnetrones, synchrotrones, free electron
lasers, etc) could in principle be modified to work in this range.
However, these devices are still in prototype form, are not compact,
or may require cryogenic temperatures to operate. The laser technology
operates at the higher frequencies, with the wavelengths below $10\,\mu$m.
At the same time, radiation sources in the THz band are required to
study rotational energy levels in complex molecules, oscillations
in solid crystals, etc \cite{spectra}. If the THz field is strong
enough (above $100\,$MV/m), nonlinear interaction with solid state
materials becomes possible including excitation of a diverse zoo of
oscillatory degrees of freedom (spin waves, phonons, magnons, excitons,
etc) \cite{Spin,Spin1,Spin2}. New physics related to control of nonequilibrium
processes in solid state, initiation of surface chemical reactions,
security, location, etc. require powerful sources of narrow band THz
radiation with a tunable central frequency \cite{Surface}. The terahertz
radiation also has a significant potential in medical diagnosis and
treatment because its frequency range corresponds to the characteristic
energy of biomolecular motion. Advantageously, terahertz-specific
low photon energy does not cause the ionization of biomolecules \cite{Medicin}

Recently, there was a lot of activity on laser- and accelerator-based
schemes of THz generation, which lead to creation of powerful THz
sources \cite{AcceleratorTHZ}. Most of these generate just a single
period of high amplitude THz emission. It is still a challenge to
create a narrowband THz source with the pulse energy beyond $1\,\mu$J.
Presently, the most powerful sources of THz radiation (1-10~MW, $10\,\mu$J
energy per pulse) are free electron lasers \cite{FEL_THz} which are
expensive and not compact.

One of the possibilities to generate a narrowband THz radiation is
to exploit plasma oscillations - or wakefields - excited by a relativistic
driver. The driver can be a short pulse laser or short bunch of charged
particles. When the driver propagates through plasma, it displaces
the plasma electrons from their equilibrium positions. This is accomplished
either by the laser ponderomotive force, or by the transverse fields
of the charged particles bunch. The plasma electrons continue to oscillate
behind the driver at the local plasma frequency. This plasma wave
is called the wakefield. The wakefield phase velocity $v_{{\rm ph}}$
simply equals the group velocity $v_{{\rm g}}$ of the driver. The
wakefield is an electrostatic plasma oscillation that normally does
not couple to electromagnetic waves. 

Yet, it is possible to cause the plasma waves to radiate. One of the
options is to collide two plasma wakefields by using counter-propagating
drivers \cite{Gorbunov,PlasmaDipole}. In the year 1958 Ginzburg and
Zheleznyakov \cite{Ginsburg} first guessed that the radioemission
from colliding electrostatic plasma waves is responsible for solar
radio bursts. Tsytovich developed a kinetic theory of nonlinear waves
coupling in plasmas \cite{Tsytovitsch1966}. Later, this mechanism
was widely accepted in astrophysics \cite{Smith,Willes,Melrose}.
It is also known that wakes generated by a laser pulse at an angle
to plasma density gradients can emit broadband radiation via mode
conversion \cite{Sheng}. This mechanism corresponds to the inverse
resonant absorption. Another known way to cause the wakefields radiating
is to apply an external magnetic field \cite{Bfield}. 

In this paper, we suggest a different coupling mechanism between the
electrostatic wakefield and electromagnetic radiation. The non-linear
wakefield current contains a non-vanishing curl. When the plasma has
a positive density gradient along the driver propagation direction,
the wake phase velocity is continuously increasing with time after
the driver has passed over. At some point, the wake phase velocity
becomes superluminal and the non-linear curl of the plasma wave current
can efficiently couple to electromagnetic modes. The wake phase velocity
can remain above the vacuum speed of light for many plasma periods.
This time is defined by the plasma density gradient scale length and
can be sufficiently long to radiate nearly all energy stored in the
wake field. Thus, an efficient regime of energy conversion from the
driver to the plasma wakefield and then to THz radiation can be found.

In the following, we assume the plasma wakefield is generated by a
short pulse laser, although a similar formalism can be applied to
charged particles bunches. We assume the plasma is tenuous, $\omega_{p}\ll\omega_{0}$,
where $\omega_{0}=2\pi c/\lambda_{0}$ is the laser frequency with
$\lambda_{0}$ being the laser wavelength, $\omega_{p}=\sqrt{4\pi e^{2}n/m}$
is the plasma frequency, $n$ is the background electron density,
$m$ is the electron mass and $e$ is the elementary charge. The laser
pulse with the dimensionless amplitude $a=\sqrt{0.74\,I/I_{18}}\lambda_{0}/\mu$m
excites a (quasi-)linear wakefield with the potential $\Phi$ \cite{Esarey}

\begin{equation}
\left(\frac{\partial^{2}}{\partial t^{2}}+\omega_{p}^{2}(z)\right)\Phi(t,z)=\omega_{p}^{2}(z)\frac{mc^{2}}{e}a^{2}\left(t,z\right).\label{eq:Wakefield-1}
\end{equation}
Here, $I$ is the laser intensity, $I_{18}=10^{18}\,{\rm W/cm^{2}}$,
$a\left(t,z\right)=a\left(t-\int^{z}v_{{\rm g}}^{-1}(\zeta)d\zeta\right)$,
and $v_{{\rm g}}(z)\approx c\left(1-\omega_{p}^{2}(z)/2\omega_{0}^{2}\right)$
is the laser group velocity. We assume the laser pulse is weakly relativstic,
$a\ll1$ and the density scale $L=\omega_{p}\left(d\omega_{p}/dz\right)^{-1}$
is very long, $\omega_{p}L/c\gg1$. The solution of this linear equation
is a harmonic oscillation

\begin{figure}
\includegraphics[width=1\columnwidth]{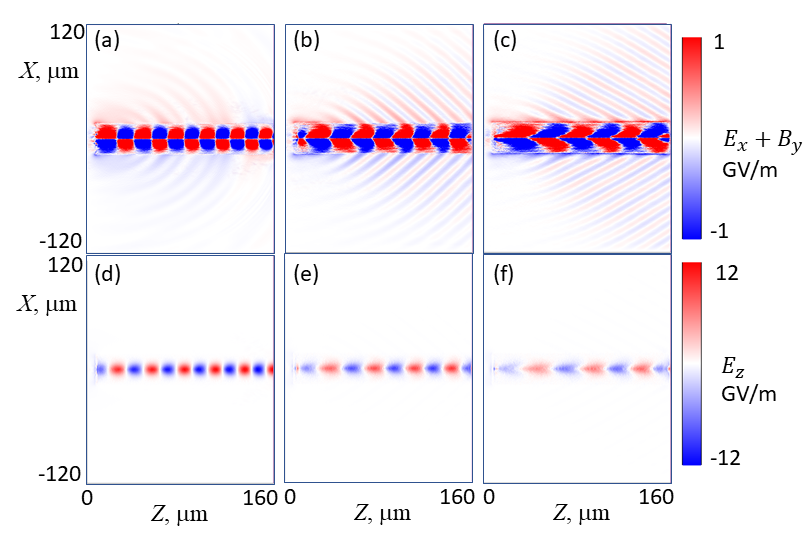}

\caption{(a)-(c) Snapshots of THz emission from a laser-plasma wakefield at
the times $t_{1,2,3}=650{\rm fs,}1450{\rm fs,}2000{\rm fs}.$ (d)-(f)
Evolution of the tongitudinal field $E_{z}$. The effective wavelength
of the wake increases with time. The angle of THz emission changes
correspondingly. \label{fig:THz-emission2D}}
\end{figure}

\begin{equation}
\Phi(t,z,{\bf r}_{\perp})=\Phi_{0}({\bf r}_{\perp})\exp\left({\bf i}\varphi(t,z)\right)+c.c.\label{eq:Wakefield}
\end{equation}
with the wake phase $\varphi(t,z)=-\omega_{p}(z)\left(t-\int^{z}v_{{\rm g}}^{-1}(\zeta)d\zeta\right)$.
The wake phase velocity simply equals the driver group velocity $v_{{\rm g}}(z)\approx c$
in the tenuous plasma. The full formula gives only a small correction,
while complicates the analytics. For this reason, we omit the small
difference and assume $v_{{\rm g}}=c$ in the following formulas.

At larger amplitudes, the wakefield becomes anharmonic. In the 1D
case, its potential satisfies the nonlinear oscillator equation \cite{Sprangle1990}

\begin{align}
\frac{d^{2}\Phi}{d\varphi^{2}} & =-\frac{mc^{2}}{e}\frac{1}{2}\left[1-\frac{1}{\left(1+\frac{e\Phi}{mc^{2}}\right)^{2}}\right]\label{eq:Wakefield-Nonlinear}\\
= & -\frac{mc^{2}}{e}\left[\frac{e\Phi}{mc^{2}}-\frac{3}{2}\left(\frac{e\Phi}{mc^{2}}\right)^{2}+2\left(\frac{e\Phi}{mc^{2}}\right)^{3}+...\right]\text{.}\nonumber 
\end{align}
The quasi-linear solution contains higher harmonics of the plasma
frequency $\omega_{l}=l\omega_{p}$, where $l$ is the harmonic number.

Our ansatz for the wake phase is $\varphi(t,z)=-\omega_{p}(z)\tau$,
where $\tau=t-z/c$ is the fast time. The wake phase frequency is
then $\omega=-\partial\varphi/\partial t=\omega_{p}(z)$ and the longitudinal
wavenumber is $k_{z}=\partial\varphi/\partial z=\omega_{p}(z)/c-\tau\partial\omega_{p}(z)/\partial z$.
The wake phase velocity along the driver propagation direction is
not constant anymore, but changes with time after the driver passage

\begin{equation}
v_{{\rm ph_{z}}}=\frac{\omega}{k_{z}}=\frac{c}{1-c\tau/L}.\label{eq:Vph}
\end{equation}
The phase velocity \eqref{eq:Vph} stays superluminal for times 

\begin{equation}
0<\tau<2\frac{L}{c}\label{eq:super}
\end{equation}
Within the period of time \eqref{eq:super}, the wakefield is resonant
with electromagnetic waves and radiates. Mention, the phase velocity
passes through a singlularity at $\tau=L/c$, then reverses and becomes
negative afterwards. It has been recently shown that this reversal
can be used to generate high-power broadband Cherenkov signal \cite{Kalmyk2021}.
Here, instead we discuss a narrow band emission at plasma frequency
harmonics.

The radiation source is obtained from the wave equation, which we
write on the magnetic field ${\bf B}$:

\begin{equation}
-\frac{\partial^{2}}{\partial t^{2}}{\bf B}+c^{2}\nabla^{2}{\bf B}-\omega_{p}^{2}{\bf B}=-4\pi c{\bf \nabla\times j}_{{\rm wake}},\label{eq:WaveEq}
\end{equation}
where ${\bf j}_{{\rm wake}}=-e(n+\delta n){\bf v}_{{\rm wake}}$ is
the current density generated by the wakefield, $n$ is the background
electron density, $\delta n$ is the density perturbation due to the
plasma wave, and ${\bf v}_{{\rm wake}}$ is the corresponding electron
velocity. The non-vanishing curl of the wake current appears due to
the nonlinear term $-e\delta n{\bf v}_{{\rm wake}}$ that causes the
mode mixing. We Fourier-transform the wakefield potential in plasma
frequency harmonics $\Phi=\int\sum_{l}\Phi_{l{\bf k}_{\perp}}\exp\left(-{\bf i}l\omega_{p}t+{\bf i}lk_{z}z+{\bf i}{\bf k_{\perp}{\bf r}_{\perp}}\right)d{\bf k_{\perp}}$,
with $|l|$ being the harmonic number, and obtain the radiation source
(compare Eq. (12.11) in \cite{Tsytovitsch1966} and Eq. (1) in \cite{Willes}):

\begin{align}
{\bf R}_{l,{\bf k_{\perp}}} & =\frac{{\bf i}ec}{2m\omega_{p}}\sum_{l_{1},l_{2}=-\infty}^{\infty}\iint\delta_{l_{1}+l_{2},l}\delta\left({\bf k_{\perp}-k_{1\perp}-k_{2\perp}}\right)\label{eq:jSource}\\
 & \left(\frac{k_{1}^{2}}{l_{2}}-\frac{k_{2}^{2}}{l_{1}}\right){\bf k_{1}}\times{\bf k_{2}}\Phi_{l_{1}{\bf k_{1\perp}}}\Phi_{l_{2}{\bf k_{2\perp}}}{\rm d}{\bf k_{1\perp}}{\rm d}{\bf k_{2\perp}}\nonumber 
\end{align}
The radiated frequency is $\omega=(l_{1}+l_{2})\omega_{p}$ and the
wave vector ${\bf k}={\bf k_{1}}+{\bf k_{2}}$ have to satisfy the
electromagnetic dispersion relation $\omega^{2}=\omega_{p}^{2}+c^{2}k^{2}$.
The source \eqref{eq:jSource} is proportional to the wake field potenital
squared. The wakefield amplitude in turn scales linearly with the
laser intensity $I_{L}$. Assuming the wavenumbers scale with $k\sim\omega_{p}/c$,
we obtain that for weakly relativistic wakefields, the power of emitted
THz radiation at the second harmonic scales as $P_{{\rm THz}}\sim|{\bf R}_{\omega,{\bf k}}|^{2}\sim n^{2}I_{{\rm L}}^{4}$.
For higher harmonics with $|l|\ge4$, the simple formula \eqref{eq:jSource}
requires relativistic corrections, which will be published elsewhere.

The formula \eqref{eq:jSource} tells us further that plane plasma
waves do not radiate. The radiation requires presence of non-collinear
wave vectors ${\bf k_{1}}$ and ${\bf k_{2}}$ with $k_{1}^{2}\neq k_{2}^{2}$
in the wave. Thus, a transversely symmetric wakefield never radiates
exactly forward. 

The emission direction can be obtained from the resonance condition.
The angle of emission $\theta_{p}$ inside the plasma column is $\cos\theta_{p}=\left(1-c\tau/L\right)/\sqrt{1-\omega_{p}^{2}/\omega^{2}}$.
This angle depends not only on the delay $\tau$ after the driver
has passed, but also on the radiated frequency $\omega$. However,
when the radiation leaves the plasma column and passes the lateral
plasma-vacuum boundary, it undergoes an additional refraction, so
that the observed angle of emission $\theta_{{\rm vacuum}}$ does
not depend anymore on the frequency and all harmonics exit plasma
at the same angle

\begin{equation}
\cos\theta_{{\rm vacuum}}=1-\frac{c\tau}{L}\label{eq:Resonance}
\end{equation}
The expression \eqref{eq:Resonance} defines the angle of emission
as a function of the delay $\tau$ after the driver passage.

We simulate the THz emission from a plasma wakefield by the fully
relativistic three-dimensional (3D) particle-in-cell code VLPL (Virtual
Laser-Plasma Laboratory) \cite{VLPL}. In the simulation, the background
gas is assumed to be hydrogen (${\rm H}_{2}$) with a linear density
gradient so that the molecular density rises from $0.85\times10^{18}{\rm cm^{-3}}$to
$1.7\times10^{18}{\rm cm^{-3}}$ over the distance of $L=0.5\,{\rm mm}$.
The hydrogen is dynamicall ionized by the short pulse laser with the
intensity $I_{{\rm L}}=5\times10^{17}\,{\rm W/cm^{2}}$ and FWHM pulse
duration $T_{{\rm L}}=30\,{\rm fs}$. This pulse duration is close
to the ``resonant'' one \cite{Gorbunov1987}. The laser pulse has
the wavelength $\lambda_{0}=800\,{\rm nm}$ and is focused to a focal
spot with the FWHM diameter $14\,\mu{\rm m}$. The focal plane position
is at $z=100\,{\rm \mu m}.$ This laser pulse has $40\,{\rm mJ}$
energy. The laser pulse group velocity here is $v_{g}\approx\left(1-\omega_{p}^{2}/2\omega_{{\rm 0}}^{2}\right)c\approx0.9995c$.
The three-dimensional simulation box was $L_{x}\times L_{y}\times L_{z}=240\,{\rm \mu m}\times240\,{\rm \mu m}\times160\,{\rm \mu m}$
sampled by grid steps $h_{z}=0.14\,{\rm \mu m}$ and $h_{x}=h_{y}=0.4\,{\rm \mu m}$
the time step was $\Delta=ch_{z}$ \cite{RIP}. The boundary conditions
were open for both fields and particles on all sides of the simulation
box.

For simplicity and to save computational resources, we simulate only
the first $150\,{\rm \mu m}$ of the density ramp. As the laser pulse
propagates, it ionizes the background gas within the focal spot and
produces a plasma column. Outside of the plasma column, the gas remains
unionized. Fig.~\ref{fig:THz-emission2D} shows 2D $(z,x)$ snapshots
of the THz emission (the right-bound field $E_{x}+B_{y}$) and the
longitudinal wake field $E_{z}$ at the times $t_{1,2,3}=650{\rm fs,}1450{\rm fs,}2000{\rm fs}.$
The excited wakefield has the amplitude $e\Phi/mc^{2}\approx0.1$
immediately behind the laser pulse \cite{Esarey}. This corresponds
to the longitudinal field of $E_{z}\approx12\,{\rm GeV/m}$.

\begin{figure}
\includegraphics[width=1\columnwidth]{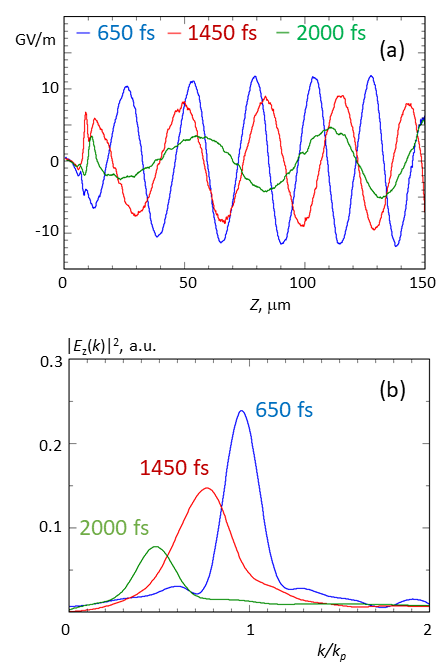}

\caption{(a) The on-axis wakefield $E_{z}$for the three times $t_{1,2,3}=650{\rm fs,}1450{\rm fs,}2000{\rm fs}.$
It is seen that the effective wavelength increases. At the same time
the wakefield amplitude drops continuously due the radiation losses.
(b) Fourier transformation of the wakefields. At the earlier time,
the wake wavenumber is centered at $k_{p}=\omega_{p}/c.$ At the later
times, the peak is shifting to smaller wavenumbers. \label{fig:THzVEGA-Wakes}}
\end{figure}

Evolution of the on-axis longitudial plasma wakefield is shown in
Fig.~\ref{fig:THzVEGA-Wakes}. One sees as the wakefield period increases
with time, thus increases the effective phase velocity. The continuous
emission of electromagnetic waves leads to the steady drop of the
wakefield amplitude.

\begin{figure}
\includegraphics[width=1\columnwidth]{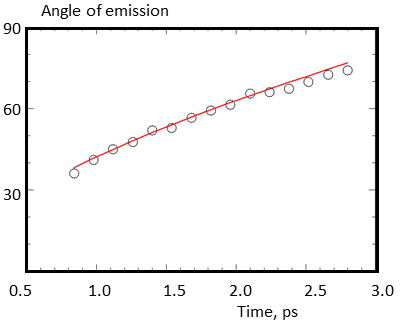}

\caption{Angle of THz emission as a function of time. Circles mark the simulation
results. The solid line is given by the formula \eqref{eq:Resonance}.
\label{fig:THz-theta}}
\end{figure}

Simultaneously, one sees as the plasma wakefield emits THz radiation
at an angle to the laser propagation direction. The angle of emission
$\theta$ can be estimated by the inclination of wave phase fronts
outside of the plasma column. This angle increases with time as predicted
by Eq. \eqref{eq:Resonance}. Fig.~\ref{fig:THz-theta} shows the
angle of emission observed in the 3D PIC simulation (circles) as a
function of time. The solid line gives the formula \eqref{eq:Resonance}. 

\begin{figure}
\includegraphics[width=1\columnwidth]{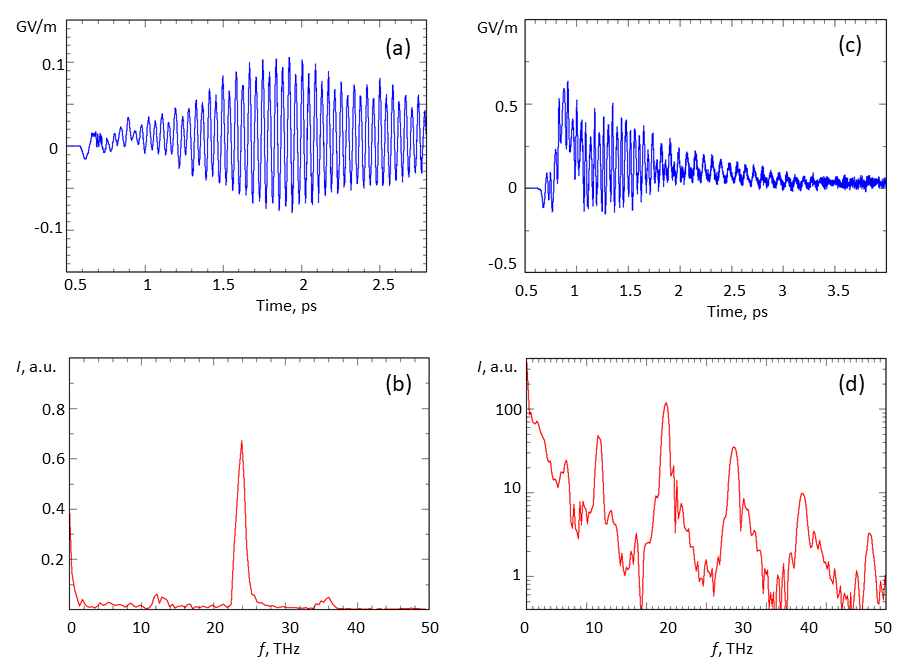}

\caption{THz signal and its spectra recorded at the right boundary the radial
position $r=100\,{\rm \mu m}$ outside of the plasma column (a)-(b)
for the laser intensity $I_{{\rm L}}=5\times10^{17}\,{\rm W/cm^{2}}$
and focal spot diameter $14\,{\rm \mu m}$; (c)-(d) for the laser
intensity $I_{{\rm L}}=2\times10^{18}\,{\rm W/cm^{2}}$ and focal
spot diameter $7\,{\rm \mu m}$. Radiation emission at the harmonics
of plasma frequency $f_{p}\approx12\,{\rm THz}$ are observed \label{fig:THzVEGA}}
\end{figure}

The radiated wave field has been recorded at the right boundary at
the point located $100\,{\rm \mu m}$ away from the laser optical
axis. The field itself and the spectrum are shown in Fig.~\ref{fig:THzVEGA}(a)-(b).
We see a nearly monochromatic signal at the second plasma frequency
$2\omega_{p}$, which lasts for couple picoseconds. The wakefield
in this case is only weakly nonlinear, so the emission at the harmonics
of plasma frequency is low. Yet, we see a weaker emission at $3\omega_{p}$
and even at the plasma frequency itself. The electromagnetic wave
at $\omega_{p}$ has an implicit transverse wavenumber $k_{\perp}$,
so it is an evanescent wave inside plasma. However, as the plasma
column radius is comparable with the plasma skin depth $c/\omega_{p}$,
the evanescent wave can reach the plasma-vacuum boundary and escape.
In addition, we see in Fig.~\ref{fig:THzVEGA}(b) a radiation at
a very low frequency. This is the Sommerfeld wave \cite{Sommerfeld,SPW}
attached to the plasma filament. The Sommerfeld wave is a slow wave
with a sub-luminal phase velocity. Because it has a very long wavelength,
it can be quasi-resonantly excited even without any density gradient. 

To highlight the THz emission at harmonics of plasma frequency, we
did another simulation with the same gas target, but focused the laser
pulse to a twice smaller spot of $7\,{\rm \mu m}$ and peak intensity
$I_{{\rm L}}=2\times10^{18}\,{\rm W/cm^{2}}$. The plasma wave produced
by the higher intensity laser pulse has a larger amplitude and radiates
more at the higher harmonics. The THz signal recorded at the same
position is shown in Fig.~\ref{fig:THzVEGA}(c). The field amplitude
exceeds $0.5\,{\rm GV/m}$ and lasts for about half a picosecond.
During this time, the wakefield radiates away nearly all its energy.
The spectrum of radiation is shown in Fig.~\ref{fig:THzVEGA}(d).
Although the second plasma harmonic still dominates, we see at least
five harmonics of the plasma frequency. Mention also the emission
at the very low frequencies, around $f\approx0$. This is again the
Sommerfeld mode (cylindrical plasma surface wave). This mode causes
the significant asymmetry of the THz field seen in Fig.~\ref{fig:THzVEGA}(c). 

In conclusion, we have shown that a plasma wakefield can efficiently
radiate electromagnetic modes, when there is a positive density gradient
along the driver propagation direction. The density scale length $L$
defines how long the wake phase velocity remains superluminal and
the coupling to the electromagnetic waves is possible. The low amplitude
linear wake emits a narrow band THz radiation at the second plasma
harmonic, $\omega=2\omega_{p}$. At higher amplitudes, the nonlinear
plasma wave emits at the higher harmonics as well. The angle of emission
is defined by the delay $\tau$ behind the driver and the density
scale length $L$. The intensity of emission scales as the wake amplitude
to the fourth power. Because there is no other dissipation mechanism
in tenuous plasmas, one can adjust the density gradient so that most
of the wake energy is depleted by the THz radiation. In this work,
we explain the physics of interaction, provide simple estimates and
show results of full 3D PIC simulations. Accurate analytical formulas
for typical Gaussian laser pulses will be presented elsewhere.

This work has been supported in parts by the Deutsche Forschungsgemeinschaft,
BMBF (Germany) and by Ministry of Science and Higher Education of
the Russian Federation (agreement No. 075-15-2020-906., Center of
Excellence \textquotedblleft Center of Photonics\textquotedblright ).

\end{document}